\documentclass[twocolumn]{aastex63}

\newcommand{\msun}{\rm M_{\odot}}

\shorttitle{Hadronic Emissions from MADs}
\shortauthors{Kimura and Toma}

\begin{document}
\title{Hadronic High-energy Emission from Magnetically Arrested Disks in Radio Galaxies}

\correspondingauthor{Shigeo S. Kimura}
\email{shigeo@astr.tohoku.ac.jp}

\author[0000-0003-2579-7266]{Shigeo S. Kimura}
\altaffiliation{JSPS Fellow}
\affiliation{Frontier Research Institute for Interdisciplinary Sciences, Tohoku University, Sendai 980-8578, Japan}
\affiliation{Astronomical Institute, Graduate School of Science, Tohoku University, Sendai 980-8578, Japan}

\author[0000-0002-7114-6010]{Kenji Toma}
\affiliation{Frontier Research Institute for Interdisciplinary Sciences, Tohoku University, Sendai 980-8578, Japan}
\affiliation{Astronomical Institute, Graduate School of Science, Tohoku University, Sendai 980-8578, Japan}




\begin{abstract}
We propose a novel interpretation that gamma-rays from nearby radio galaxies are hadronic emission from magnetically arrested disks (MADs) around central black holes (BHs). The magnetic energy in MADs is higher than the thermal energy of the accreting plasma, where the magnetic reconnection or turbulence may efficiently accelerate non-thermal protons. They emit gamma-rays via hadronic processes, which can account for the observed gamma-rays for M87 and NGC 315. Non-thermal electrons are also accelerated with protons and produce MeV gamma-rays, which is useful to test our model by proposed MeV satellites. The hadronic emission from the MADs may significantly contribute to the GeV gamma-ray background and produce the multi-PeV neutrino background detectable by IceCube-Gen2. In addition, gamma-rays from MADs provide electron-positron pairs through two-photon pair production at the BH magnetosphere. These pairs can screen the vacuum gap, which affects high-energy emission and jet-launching mechanisms in radio galaxies.
\end{abstract}



%
\section{Introduction}

Radio-loud active galactic nuclei (AGNs) have prominent relativistic jets. They are less common than radio-quiet AGNs, but have strong influence on evolutions of host galaxies and galaxy clusters owing to their powerful jets. Radio-loud AGNs are classified into two sub-classes. One is blazars whose jets are directed to the Earth. They emit powerful broadband radiations, from radio to TeV gamma-rays, owing to the relativistic beaming effect \citep[see][for a recent review]{2019Galax...7...20B}.  The other is radio galaxies whose jets are misaligned to the Earth. Despite that their jets are seen from off-axis, some of the radio galaxies are observed from radio to GeV-TeV gamma-rays. The emission mechanism and production site of the gamma-rays are still controversial \citep[see][for a review]{2018Galax...6..116R,2019Galax...8....1H}. 

Regarding M87, a nearby radio galaxy  whose jet and central region are investigated in great detail \citep[e.g.,][]{2011Natur.477..185H,2019ApJ...875L...1E,2019ApJ...875L...2E,2019ApJ...875L...3E,2019ApJ...875L...4E,2019ApJ...875L...5E,2019ApJ...875L...6E,2019ApJ...887..147P}, no model can satisfactorily explain the origin of gamma-rays. For instance, the leptonic jet models require the magnetic field of a few mG \citep{2009ApJ...707...55A,2020MNRAS.492.5354M}, which is much weaker than the value estimated by the radio observation \citep{2013ApJ...775...70H,2015ApJ...803...30K,2019MNRAS.489.1633L}. The magnetic field can be strong enough for the hadronic jet models, but they demand a jet power much higher than the estimated mass accretion power \citep{2011A&A...531A..30R} or a high beaming factor despite that the jet is off-axis \citep{2020MNRAS.492.5354M}. A hybrid model may overcome the shortcomings above \citep{2016ApJ...830...81F}. However, it requires the emission region as compact as the central black hole (BH), which is too small for the standard dissipation scenario, such as internal shocks or collimation shocks.  The large-scale ($\gtrsim$ kpc) jet models are also proposed \citep{2003ApJ...597..186S,2011MNRAS.415..133H}, but they are too faint for standard parameters and at odds with the observed time variability (weeks to months) in GeV-TeV bands \citep{hess06m87,2014ApJ...788..165H,2019A&A...623A...2A}\footnote{Although magnetic reconnection models at a large scale, where smaller plasmoids moving with a relativistic speed emit high-energy gamma-rays, may help reducing the tensions \citep[e.g.,][]{2010MNRAS.402.1649G,2016MNRAS.462.3325P}, such models require a high magnetization parameter at the large scale, which is unlikely due to the conversion of magnetic energy to bulk kinetic or thermal energies at smaller scales as indicated by various observations.}.  The BH magnetosphere models in which a vacuum gap accelerates electron-positron pairs may be feasible for TeV gamma-rays, but reproducing the GeV gamma-ray data is challenging due to a hard photon spectrum and very-high maximum energy of electrons \citep{2011ApJ...730..123L,2016ApJ...818...50H,2020ApJ...902...80K}.

  \begin{figure}
   \begin{center}
    \includegraphics[width=\linewidth]{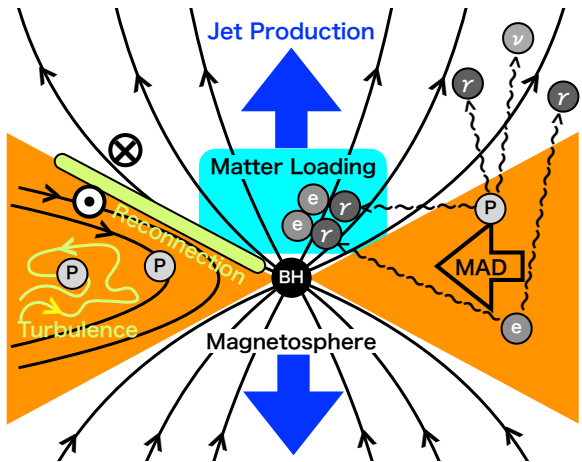}
    \caption{Schematic picture of our model. Protons are accelerated in the MAD through reconnection or turbulence, leading to hadronic gamma-ray and neutrino emissions. The gamma-rays interact with lower-energy photons emitted by thermal electrons, efficiently creating the electron-positron pairs in the magnetosphere.   }
    \label{fig:schematic}
   \end{center}
  \end{figure}

In this paper, we propose hadronic processes in magnetically arrested disks \citep[MADs;][]{NIA03a} as an alternative gamma-ray emission mechanism. Radio galaxies likely host MADs because they can efficiently launch relativistic jets \citep{TNM11a,MTB12a,SNP13a,2019MNRAS.486.2873C,2019ApJ...875L...5E,2019ApJS..243...26P} by the Blandford-Znajek mechanism \citep{BZ77a,Kom04a,TT16a}. 
The other accretion mode, the standard and normal evolution (SANE), produces weaker jets \citep{2019ApJ...875L...5E}, and these two accretion modes may cause the observed dichotomy of radio-loud and radio-quiet AGNs. The estimate of magnetic fluxes by radio observations also supports existence of MADs in radio galaxies \citep{2014Natur.510..126Z,2015MNRAS.451..927Z}.  
MADs dissipate their magnetic energies through plasma processes, such as magnetic reconnection \citep{BOP18a,2020ApJ...900..100R}, and non-thermal particles are efficiently accelerated by reconnection \citep{2001ApJ...562L..63Z,Hos12a,2014ApJ...783L..21S,2018MNRAS.473.4840W} and/or turbulence \citep{lyn+14,KTS16a,2018PhRvL.121y5101C,2019MNRAS.485..163K}, leading to gamma-ray emission via hadronic and leptonic processes (see Figure \ref{fig:schematic}). 

Hadronic emission from the accretion flows were previously discussed as the emission mechanisms of soft gamma-rays \citep{mnk97,OM03a,nxs13}, TeV gamma-rays \citep{2019ApJ...879....6R,2019arXiv190305249R,2018arXiv181102812R}, and TeV-PeV neutrinos \citep{kmt15,KG16a,2019PhRvD.100h3014K,2019ApJ...886..114H,Murase:2019vdl,Kimura:2020thg,Murase:2020lnu}. 
Multi-wavelength and multi-messenger observations also provide direct hints of non-thermal activities in accretion flows, such as Sgr A* flares in infrared and X-ray bands \citep{2010RvMP...82.3121G}, detection of GeV gamma-rays from radio-quiet AGNs \citep{WNX15a,2020ApJ...892..105A}, and a neutrino hotspot coincident with a radio-quiet AGN \citep{Aartsen:2019fau}. However, the accretion flows have not been examined as the GeV gamma-ray emission sites, especially for gamma-ray loud radio galaxies.  

In addition, high-energy phenomena in accretion flows may play an essential role on injecting particles in relativistic jets. Because of the centrifugal force and the magnetic field barrier, the accreting matter cannot enter the polar region of the BH, which results in lack of the mass supply, leading to a continuous density decrease \citep{2004ApJ...611..977M,TNM11a}. The polar region of the BH, or the funnel, is the launching point of the relativistic jets, and hence, a steady jet production needs mass and charge loading mechanisms. Also, vacuum gaps may open at the extremely low density environment, where the electron-positron pairs are accelerated and emit very high-energy gamma-rays \citep{2016ApJ...818...50H,2015ApJ...809...97B,2020ApJ...902...80K}.  Hadronic interactions create neutral particles, e.g., neutrons and gamma-rays, and they can penetrate the magnetic field barrier, enabling mass and charge loading to the funnel. In the previous studies that consider only MeV photons by thermal electrons,  the amount of pairs is marginal to screen the vacuum gap  \citep{2011ApJ...730..123L,2011ApJ...735....9M}. GeV-TeV gamma-rays can enlarge the pair density in the funnel, which screen the vacuum gap. As for the past studies for mass loading through the neutron decay, the cosmic-ray loading parameter is treated as a free parameter \citep{tt12,ktt14}. Our MAD model enables us to calibrate it using gamma-ray observations. 

This paper is organized as follows. We develop a MAD model that can reproduce gamma-ray observations, which is described in Section \ref{sec:model}. We show the broad-band spectra for the nearby radio galaxies, M87 and NGC 315, which are in low accretion states and detected by {\it Fermi}. In Section \ref{sec:diffuse}, we estimate the contribution by MADs to the diffuse high-energy gamma-ray and neutrino backgrounds. We discuss the charge and mass loading into the funnel in Section \ref{sec:load}, and discussions on other topics are written in Section \ref{sec:discussion}. Section \ref{sec:summary} summarizes our results.
 We use the notation of $Q_X=Q/10^X$ in cgs unit, except for the BH mass, $M$ ($M_9=M/[10^9~\msun])$.

%


\section{MAD Model}\label{sec:model}

\begin{table*}[t]
\begin{center}
\caption{List of model parameters and physical quantities. The references for BH masses and distances are \citet{2019ApJ...875L...1E} for M87 and \citet{2018A&A...616A.152S} for NGC 315.}\label{tab:param}
Parameters of our model
 \begin{tabular}{|cccccccc|ccc|}
\hline
  $\alpha$ & $\beta$& $\mathcal R$ & $\epsilon_{\rm dis}$ & $\eta$ & $\epsilon_{\rm NT}$ & $s_{\rm inj}$ & $\epsilon_j$ &  $M~[10^9~\msun]$ & $d_L~[(\rm Mpc)]$ & $\dot{m}~[10^{-4}]$\\
 & & & & & & & & (M87, NGC 315) & (M87, NGC 315) & (M87, NGC 315) \\
\hline
  0.3 & 0.1 & 10 & 0.15 & 5 &  0.33 & 1.3 & 1.0 & (6.3, 1.7) & (17, 65) & (0.5, 4.0) \\
\hline
 \end{tabular}

Physical quantities
\begin{tabular}{|c|cccc|cc|ccc|}
\hline
  & $B$ & $T_e$  & $Q_p/Q_e$ & $L_{\gamma,\rm thrml}$ & $L_p$& $E_{p,\rm cut}$  & $B_h$ & $n_{\rm GJ}$ & $\log(n_\pm/n_{\rm GJ})$ \\
  & [G] & [MeV] &  &  [$10^{41}\rm~erg~s^{-1}$] &[$10^{41}\rm~erg~s^{-1}$] & [EeV] &   [kG] & [$10^{-5}\rm~cm^{-3}$] & \\
\hline
M87 & 18 & 2.2 & 12 & 3.4 & 20 &  8.1 & 0.31 & 2.8 & 2.08\\
NGC315 & 98 & 1.3 & 13 & 6.4 & 43 &  5.4 & 1.70 & 56.0 & 3.62\\
\hline
\end{tabular}

\end{center}
\end{table*}

  \begin{figure*}[t]
   \begin{center}
    \includegraphics[width=\linewidth]{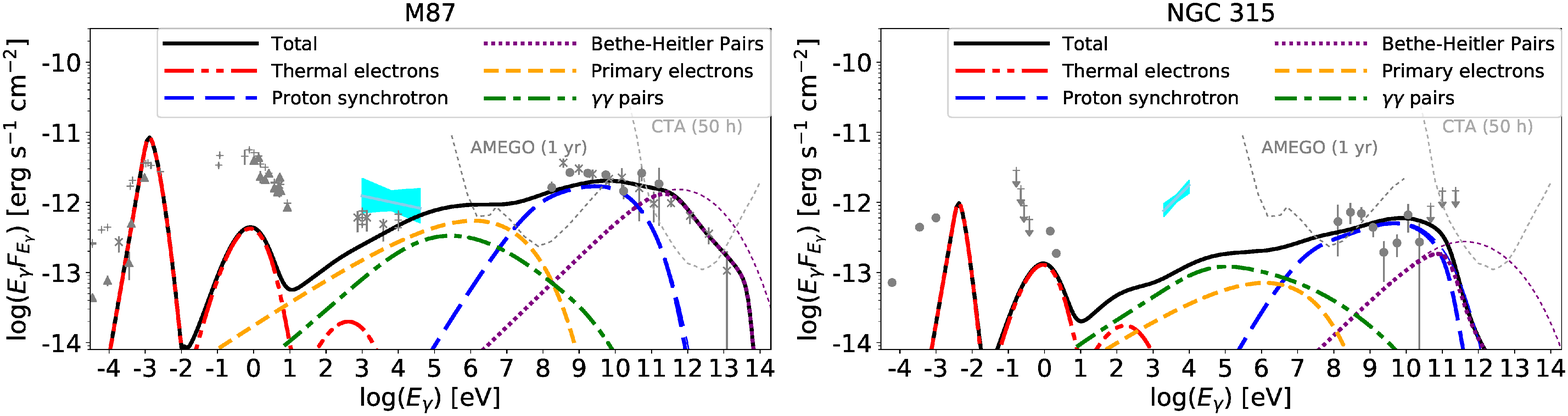}
    \caption{Broadband photon spectra for M87 (left) and NGC 315 (right). 
The thick lines are the observed flux on Earth, and the thin lines are the intrinsic spectra before the attenuation. The thin-dotted lines are the sensitivity for CTA \citep{2019scta.book.....C} and AMEGO \citep{2017ICRC...35..798M}. Data points are taken from \citet{2020MNRAS.492.5354M,2016MNRAS.457.3801P,2017ApJ...849L..17W,2019A&A...623A...2A} for M87 and from \citet{2020MNRAS.492.4120D} for NGC 315.}
    \label{fig:spe}
   \end{center}
  \end{figure*}

  \begin{figure*}[t]
   \begin{center}
    \includegraphics[width=\linewidth]{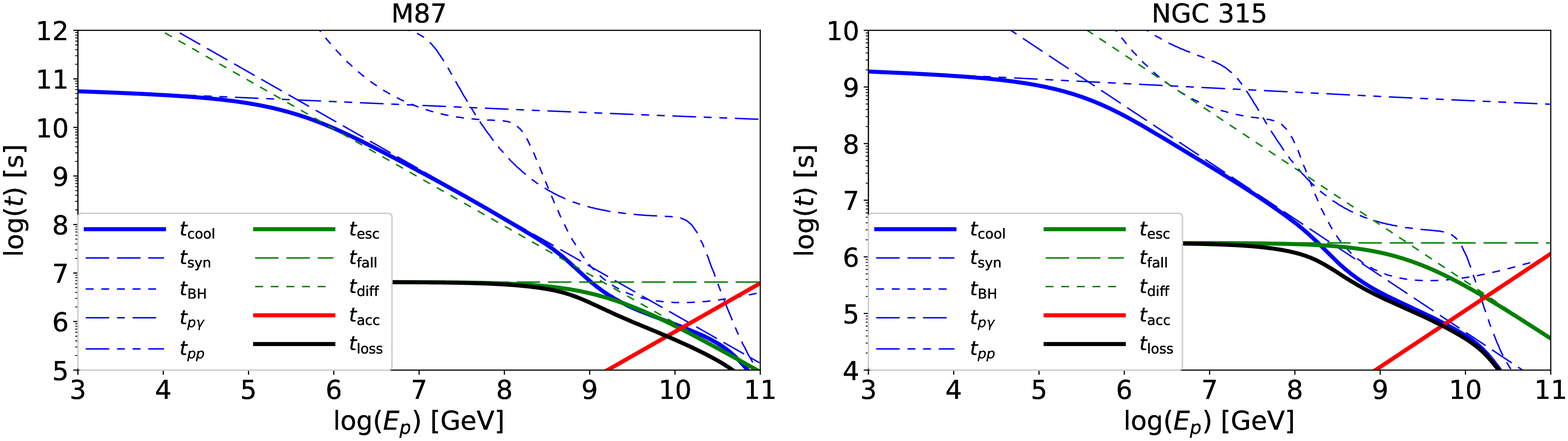}
    \caption{Cooling (blue), escape (green), loss (black), and acceleration (red) timescales as a function of the proton energy for M87 (left) and NGC 315 (right). $t_{\rm syn}$, $t_{\rm BH}$, $t_{p\gamma}$, and $t_{pp}$ are the synchrotron, Bethe-Heitler, photomeson production, and proton-proton cooling timescales, respectively.  }
    \label{fig:times}
   \end{center}
  \end{figure*}

  We calculate the photon spectra from MADs using one-zone and steady-state approximations. Since the analytic prescription for physical quantities in MADs has not been established yet, we use the prescription for SANE-mode radiatively inefficient accretion flows (RIAFs) similar to that in \citet{2019PhRvD.100h3014K}, adjusting the reference parameter set suitable to MADs. We consider an accreting plasma onto a supermassive BH of mass $M$. The mass accretion rate, $\dot{M}$, and size of the plasma, $R$, are normalized by the Eddington rate and gravitational radius, $R=\mathcal{R}R_G=\mathcal{R}GM/c^2$ and $\dot{M}c^2=\dot{m}L_{\rm Edd}$ (without including the radiation efficiency factor), respectively. The radial velocity  and magnetic field in the MAD are analytically estimated to be \citep{2019PhRvD.100h3014K}
\begin{eqnarray*}
& V_R& \approx \frac12\alpha V_K\simeq1.5\times10^9\mathcal{R}_1^{-1/2}\alpha_{-0.5}\rm~cm~s^{-1}\\
& B& \approx \sqrt{\frac{8\pi \rho C_s^2}{\beta}}\simeq62\mathcal{R}_1^{-5/4}\dot{m}_{-4}^{1/2}M_9^{-1/2}\alpha_{-0.5}^{-1/2}\beta_{-1}^{-1/2}\rm~G,
\end{eqnarray*}
where $\alpha$ is the viscous parameter \citep{ss73}, $\beta$ is the plasma beta, $V_K=\sqrt{GM/R}$ is the Keplerian velocity, $C_s\approx V_K/2$ is the sound speed, $\rho\approx \dot M/(4\pi RHV_R)$ is the mass density, and $H\sim (C_s/V_K)R$ is the scale height.
These are in rough agreement with  magnetohydrodynamic (MHD) simulations \citep{MM03a,2012MNRAS.426.3241N,2019MNRAS.485..163K,2019MNRAS.486.2873C}. Because of the strong magnetic field, the angular momentum transport in MADs is likely more efficient than in SANEs. In this study, we use $\alpha=0.3$ and $\beta=0.1$, which are appropriate for MADs \citep{2012MNRAS.426.3241N,2019MNRAS.486.2873C,2019ApJ...874..168W}. 

We consider emissions from thermal electrons, non-thermal protons, primary electrons created together with protons, and secondary electron-positron pairs. First, we explain how to calculate photon spectra by thermal electrons.
The spectra for the synchrotron, bremsstrahlung, and Comptonization processes by the thermal electrons are calculated by the methods given in \citet{kmt15}. The electron temperature is determined such that the resulting photon luminosity is equal to the electron heating rate. The electron heating mechanism in hot accretion flows has yet to be established \citep[e.g.,][]{2015ApJ...800...88S,2019PhRvL.122e5101Z,2019PNAS..116..771K}. We utilize the formalism by \citet{2018ApJ...868L..18H}, where the ratio of heating rate for electrons to protons is given by
\begin{equation}
\frac{Q_e}{Q_p}\approx\left(\frac{m_eT_e}{m_pT_p}\right)^{1/4} .
\end{equation}
Then, the photon luminosity by the thermal electrons is estimated to be 
\begin{equation}
 L_{\gamma,\rm thrml}\approx\frac{Q_e}{Q_p}(1-\epsilon_{\rm NT})\epsilon_{\rm dis}\dot{m}L_{\rm Edd}, 
\end{equation}
where $\epsilon_{\rm NT}$ is the energy fraction of non-thermal particle production to dissipation and $\epsilon_{\rm dis}$ is the energy fraction of the dissipation to accretion.

The photon spectra by the thermal electrons for M87 and NGC 315 are shown in Figure \ref{fig:spe}, and the parameters and resulting quantities are tabulated in Table \ref{tab:param}. The MAD heats up the thermal electrons to a few MeV, and they emit peaky signals in $\sim10^{-3}$ eV by the synchrotron radiation. These photons are important targets for the photo-hadronic processes. Although the inverse Compton scattering and bremsstrahlung create higher energy photons, they are too faint to explain the observed data, and too tenuous to work as target photons.

To obtain the non-thermal spectra for particle species $i$, we solve the steady-state transport equation:
\begin{equation}
 -\frac{d}{dE_i}\left(\frac{E_iN_{E_i}}{t_{i,\rm cool}}\right)=\dot{N}_{E_i,\rm inj}-\frac{N_{E_i}}{t_{\rm esc}},
\end{equation}
 where $N_{E_i}$ is the differential number spectrum, $t_{i,\rm cool}$ is the cooing time, $t_{\rm esc}$ is the escape time, and $\dot{N}_{E_i,\rm inj}$ is the injection terms. The analytic solution of this equation is given by 
\begin{equation}
 N_{E_i}=\frac{t_{i,\rm cool}}{E_i}\int_{E_i}^{\infty}dE'_i\dot{N}_{E'_i,\rm inj}\exp\left(-\int_{E_i}^{E'_i}\frac{t_{i,\rm cool}d\mathcal{E}_i}{t_{\rm esc}\mathcal{E}_i}\right).\label{eq:NE_i}
\end{equation} 
The escape term is common for all the components. We consider diffusion and advection (infall to the BH) as the escape processes, whose timescales are estimated to be $t_{\rm diff}\approx R^2/D_R$ and $t_{\rm fall}\approx R/V_R$, respectively, 
where $D_R\approx\eta{r_{i,L}}c/3$ is the diffusion coefficient, $r_{i,L}=E_i/(eB)$ is the Larmor radius,  $\eta r_L$ the effective mean free path, and $\eta$ is a numerical factor. The total escape time is given by $t_{\rm esc}^{-1}=t_{\rm fall}^{-1}+t_{\rm diff}^{-1}$. 

The injection terms for primary protons and electrons are written as 
\begin{equation}
 \dot{N}_{E_i,\rm inj}\approx\dot{N}_0\left(\frac{E_i}{E_{i,\rm cut}}\right)^{-s_{\rm inj}}\exp\left(-\frac{E_i}{E_{i,\rm cut}}\right),
\end{equation}
where $s_{\rm inj}$ is the injection spectral index. The normalization, $\dot{N}_0$, is determined by $L_p=\int\dot{N}_{E_p,\rm inj}E_pdE_p=\epsilon_{\rm NT}\epsilon_{\rm dis}\dot{M}c^2$ for protons and $\int\dot{N}_{E_e,\rm inj}E_edE_e=(Q_e/Q_p)\epsilon_{\rm NT}\epsilon_{\rm dis}\dot{M}c^2$ for primary electrons. $E_{i,\rm cut}$ is determined by $t_{\rm loss}=t_{\rm acc}$, where $t_{\rm acc}$ is the acceleration time and $t_{\rm loss}^{-1}=t_{\rm cool}^{-1}+t_{\rm esc}^{-1}$ is the total loss timescale. We phenomenologically write the acceleration time as 
\begin{equation}
t_{\rm acc}\approx\frac{\eta r_L}{c}\left(\frac{c}{V_A}\right)^2,
\end{equation}
where $V_A/c\approx0.71\mathcal{R}_1^{-1/2}\beta_{-1}^{-1/2}$ is the Alfv\'{e}n velocity. 

As the proton cooling process, we take into account the proton synchrotron, $pp$ inelastic collisions, photomeson production ($p+\gamma\rightarrow p+\pi$), Bethe-Heitler ($p+\gamma\rightarrow p+e^++e^-$) processes. Their cooling timescales are given in \citet{2019PhRvD.100h3014K}, in which we appropriately take into account the energy-dependent crosssection for both $pp$ \citep{2014PhRvD..90l3014K} and $p\gamma$ \citep{SG83a,CZS92a,MN06b} interactions. Figure \ref{fig:times} indicates the timescales as a function of $E_p$ for M87 (left) and NGC 315 (right), whose parameters are tabulated in Table \ref{tab:param}. For $E_p\lesssim1$ EeV, the cooling is very inefficient, while the Bethe-Heitler and synchrotron processes are efficient for higher energies. For a higher $\dot{m}$, the synchrotron cooling is more efficient owing to stronger magnetic fields. The cooling limits the particle acceleration, and the maximum attainable energies for these objects are tabulated in Table \ref{tab:param}. These energies are lower than the Hillas energy, $E_{\rm Hil}\approx eBR(V_A/c)\sim35$ EeV for M87 and 52 EeV for NGC 315.  

As the cooling process for primary electrons and secondary electron-positron pairs, we consider only synchrotron radiation, because other processes are negligible. For the pairs created by Bethe-Heitler process, the injection term of the pairs is approximated to be 
\begin{equation}
E_{\rm BH,\pm}^2\dot{N}_{E_{\rm BH,\pm},\rm inj}\approx E_p^2N_{E_p}t_{\rm BH}^{-1},
\end{equation}
where $E_{\rm BH,\pm}\approx(m_e/m_p)E_p$ is the energy of the pairs by proton-photon interactions \citep{Murase:2019vdl}. 
Gamma-rays produced by non-thermal particles are absorbed by lower energy photons through the two-photon pair production, which also produces non-thermal pairs. The optical depth by $\gamma\gamma$ pair production, $\tau_{\gamma\gamma}$, is estimated by the method in \citet{cb90}, and the resulting gamma-rays are attenuated by the factor of $F_{\rm atn}\approx(1-\exp[-\tau_{\gamma\gamma}])/\tau_{\gamma\gamma}$. The injection term by $\gamma\gamma$ pair production is approximated to be 
\begin{equation}
 E_{\gamma\gamma,\pm}^2\dot{N}_{E_{\gamma\gamma,\pm},\rm inj}\approx(1-F_{\rm atn})E_\gamma{L}_{E_\gamma},
\end{equation}
where $E_{\gamma\gamma,\pm}\approx{E}_\gamma/2$ is the energy of the pairs by two-photon interactions and $L_{E_\gamma}$ is the intrinsic differential photon luminosity by all the components. Since the injection term by $\gamma\gamma$ pair production depends on the gamma-ray spectrum, we iteratively calculate the photon and pair spectra until the solutions converge. The photons from the protons and non-thermal electron-positron pairs have little influence on the photomeson production efficiency for the range of our interest, and we ignore the feedback to the photomeson production rate.





We calculate the photon spectra by non-thermal protons, primary electrons, and secondary electron-positron pairs using the non-thermal spectra obtained by Eq. (\ref{eq:NE_i}). For all the populations, synchrotron emission is the dominant radiation process, and other ones are negligible. We calculate the synchrotron spectra using the method in \citet{2008ApJ...686..181F}, appropriately taking into account the difference between protons and electrons.

The resulting gamma-ray spectra are plotted in Figure \ref{fig:spe} (see Table \ref{tab:param} for the parameter sets).  We should regard the observational data as upper limits, since other emission regions, such as inner jets and outer accretion flows, may contribute to the observed flux. For M87, the proton synchrotron emission accounts for the {\it Fermi} data, and the secondary pairs by Bethe-Heitler process reproduce the TeV gamma-rays for the quiescent state detected by MAGIC. TeV gamma-rays from M87 are strongly variable, and the flaring activities should be attributed to jets as they correlate with the radio and/or X-ray flares \citep{hess06m87,2012ApJ...746..151A,2014ApJ...788..165H}.
The primary electrons and pairs by $\gamma\gamma$ pair production emit MeV gamma-rays, which are detectable by proposed MeV satellites, such as AMEGO \citep{2017ICRC...35..798M}, e-ASTROGAM \citep{2017ExA....44...25D} and GRAMS \citep{2020APh...114..107A}.  For NGC 315, the proton synchrotron produces the gamma-rays in the {\it Fermi} band, but the TeV gamma-rays by the Bethe-Heitler pairs are completely absorbed, and we cannot detect them even with CTA. Detecting the MeV gamma-rays by the planned satellites is also challenging, although sub-GeV photons are detectable. 
Interestingly, the resulting photon spectra are almost flat for a wide energy range, because each emission component contributes to a different energy band with a similar emission power. This is a distinctive feature from one-zone leptonic jet models that typically exhibit a ``two-hump'' spectrum.

Our parameter set, with which we can account for GeV gamma-ray data, can be appropriate for MADs, but not suitable for SANEs where $\beta\gtrsim3$ and $\epsilon_{\rm NT}\lesssim0.03$ is appropriate. First, the non-thermal proton production rate is high, $\epsilon_{\rm NT}=0.33$, which can be achieved by relativistic magnetic reconnections, i.e., plasma magnetization parameter $\sigma>1$ (see Section \ref{sec:discussion}). This condition is realized in MADs, whereas we cannot find such a strong magnetization in SANEs \citep{BOP18a}. For the stochastic acceleration process expected in SANEs, $\epsilon_{\rm NT}$ is likely limited by the energy density of turbulence, which leads to $\epsilon_{\rm NT}\lesssim0.03$ \citep{2019MNRAS.485..163K}. Second, the proton synchrotron radiation is the key process to reproduce the GeV gamma-ray data, which is efficient only in MADs. SANEs have 30-100 times higher values of $\beta$, which reduces the synchrotron emissivity by the same factor. Also, synchrotron emissivity is proportional to the proton energy. We assume efficient particle acceleration in MADs, while the stochastic acceleration expected in SANEs are slower, leading to a lower proton maximum energy. Hence, hadronic gamma-ray emission is very weak in the SANE-mode RIAFs. In Figure 3 of \citet{2019PhRvD.100h3014K}, we can see that gamma-ray fluxes from nearby low-luminosity AGNs predicted by the SANE model are much lower than the {\it Fermi} sensitivity, despite that they are much brighter than M87 and NGC 315 in X-rays.


\section{Diffuse High-energy Backgrounds}\label{sec:diffuse}
  \begin{figure}
   \begin{center}
    \includegraphics[width=\linewidth]{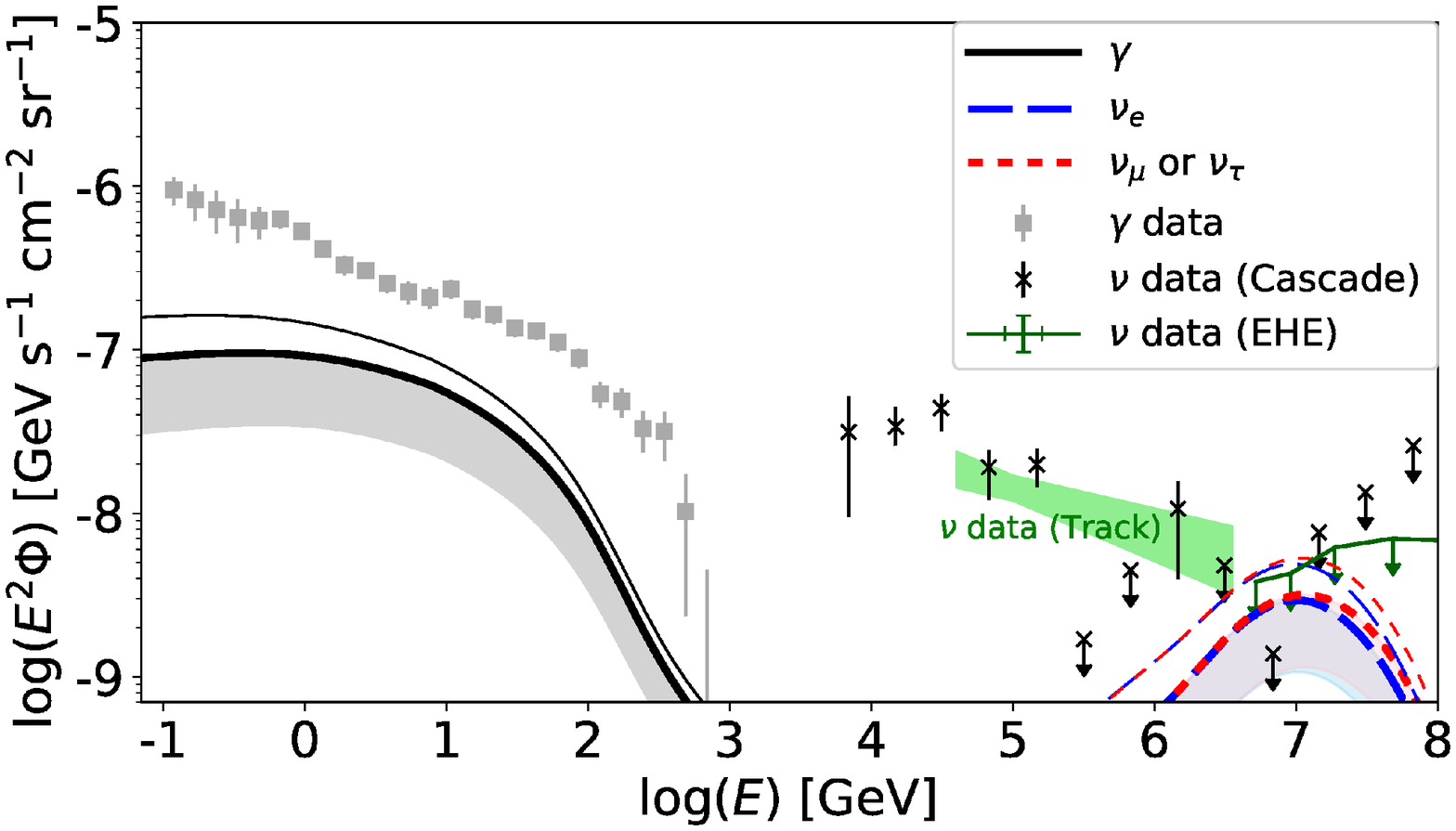}
    \includegraphics[width=\linewidth]{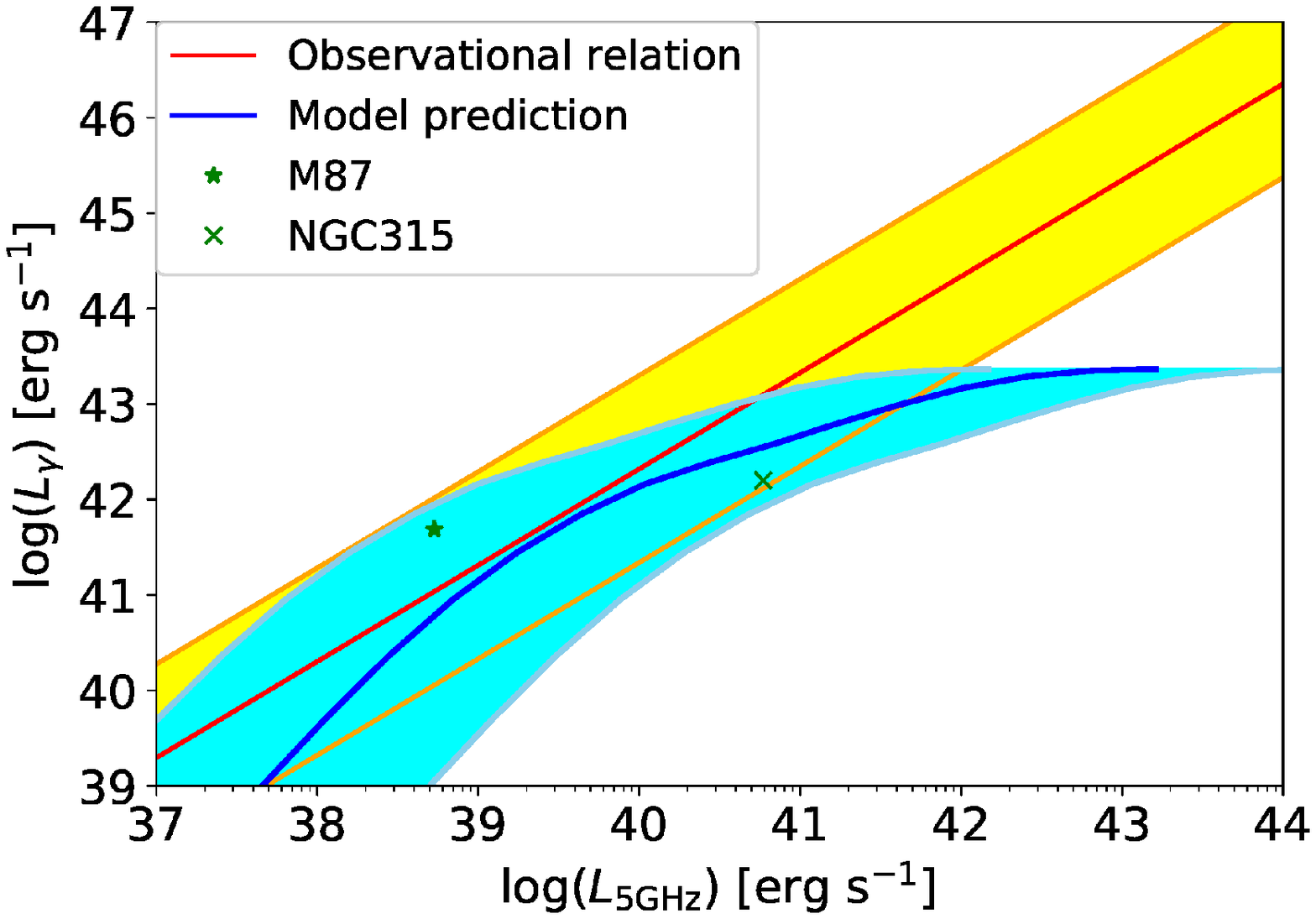}
    \caption{Top: the diffuse intensities for gamma-rays (black-solid), electron neutrinos (blue-dashed), and muon neutrinos (red-dotted). The thick and thin lines are drawn with $\epsilon_j=1$ and $\epsilon_j=0.6$, respectively. The shaded regions indicate the uncertainty by radio luminosity functions. The gamma-ray, cascade-neutrino, track neutrino, and EHE neutrino data points are from \citet{Ackermann:2014usa}, \citet{Aartsen:2020aqd}, \citet{2019ICRC...36.1017S}, and \cite{Aartsen:2018vtx}, respectively. Bottom:  Relation of 5-GHz luminosity and gamma-ray luminosity in the LAT band. The cyan and yellow regions are our model prediction and the observed correlation by \citet{2014ApJ...780..161D}, respectively.}
    \label{fig:diff}
   \end{center}
  \end{figure}

Radio galaxies are proposed as a source of the cosmic high-energy backgrounds of gamma-rays \citep{2011ApJ...733...66I,2014ApJ...780..161D,2019ApJ...879...68S} and neutrinos \citep{bec+14,2016JCAP...09..002H}. 
We here estimate the high-energy background intensities from MADs.  The $p\gamma$ neutrino spectra are obtained using method in \citet{KMM17b,KMB18a}, where the effects of pion and muon coolings and neutrino oscillation are taken into account. We evaluate the neutrino flux by $pp$ interactions using the method in \citet{kab06}, and find that they are negligible for the range of our interest.  High-energy gamma-rays from high redshifts are attenuated through interactions with the extragalactic background light (EBL), and we use optical depth for $\gamma\gamma$ attenuation in \citet{2017A&A...603A..34F}. The electromagnetic cascades initiated by EBL attenuation are negligible for the GeV gamma-ray background. 

Radio galaxies are classified into two groups based on the optical line properties. One is low-excitation radio galaxies (LERGs), which have hot accretion flows around BHs. The other is high-excitation radio galaxies (HERGs), where BHs are surrounded by optically thick, cool disks. Since our model is applicable only to the hot accretion flows, we focus on the contribution by LERGs in this study.
The local 1.4-GHz radio luminosity function of LERGs are written as $\rho_{1.4\rm GHz}=\rho_*/[(L_{1.4\rm GHz}/L_*)^{0.42}+(L_{1.4\rm GHz}/L_*)^{1.66}]$, where $L_*=1.2\times10^{41}\rm~erg~s^{-1}$ and $\rho_*=2.0\times10^{-6}\rm~Mpc^{-3}$~\citep{2014ARA&A..52..589H}\footnote{Many references for luminosity functions provide the source density in units of $\rm~Mpc^{-3}~mag^{-1}$ (equivalent to $\rm~Mpc^{-3}~\log(L)^{-1}$).  We need to multiply $\log_{10}e$ for the unit conversion from Mpc$^{-3}$ mag$^{-1}$ to Mpc$^{-3}$.} .  The LERG population shows little or no redshift evolution \citep{2016MNRAS.460....2P}, which is consistent with other low-luminosity AGN populations \citep{PMK11a,aje+14,2014ApJ...786..104U}.
1.4-GHz signals are attributed to the emission from the jet.  We use the empirical correlation, $\log(L_{j,42})=0.75\log(L_{1.4\rm GHz,40})+1.91$, to convert radio luminosity, $L_{\rm 1.4GHz}$, to the jet power, $L_j$  \citep{2010ApJ...720.1066C}\footnote{We approximately estimate the integrated radio luminosity to be $L_{\nu_0}=\nu_0P_{\nu_0}$, where $\nu_0$ is the radio frequency and $P_{\nu_0}$ is the differential luminosity in units of $\rm~erg~s^{-1}~Hz^{-1}$}. Then, we relate the jet power to the accretion rate by $L_j\sim\epsilon_j\dot{M}c^2$, where $\epsilon_j$ is the  jet production efficiency. We use $\epsilon_j=1$ as the reference value, which is supported on the simulations \citep{TNM11a,SNP13a,2019ApJ...875L...5E} and observations \citep{2011ApJ...727...39M,2015MNRAS.449..316N,2015ApJ...806...59T}. The radio galaxies of $M\simeq10^9~\msun$ mainly contribute to the radio luminosity density \citep{2005MNRAS.362...25B}, and we fix $M=10^9~\msun$ for the estimates of the diffuse intensities. Then, we convert $L_{1.4\rm GHz}$ to $\dot{m}$ and calculate neutrino and gamma-ray spectra for various $\dot{m}$ to obtain the cosmic background intensities.

The resulting high-energy neutrino and gamma-ray backgrounds are shown in the top panel of Figure \ref{fig:diff}. MADs can provide a significant contribution to the gamma-ray background for $E_\gamma\sim1-30$ GeV, up to 20 \%. The luminosity function has a break at $L_*\simeq1.2\times10^{41}\rm~erg~s^{-1}$, and LERGs with $L_{1.4\rm GHz}=L_*$, corresponding to $\dot{m}=\dot{m}_*\sim0.004$, predominantly contribute to the diffuse backgrounds. Then, the gamma-ray intensity is estimated to be 
\begin{eqnarray}
 & &E_\gamma^2\Phi_\gamma\approx\frac{c\xi_zf_{\gamma/p}f_{\rm bol}}{4\pi H_0}\rho_*\epsilon_{\rm NT}\epsilon_{\rm dis}\dot{m}_*L_{\rm Edd}\simeq7\times10^{-8}\\
&\times&M_9\left(\frac{\epsilon_{\rm NT}\epsilon_{\rm dis}}{0.05}\right)\left(\frac{\xi_z}{0.6}\right)\left(\frac{f_{\rm bol}f_{\gamma/p}}{0.1}\right){\rm~GeV~s^{-1}~cm^{-2}~sr^{-1}},\nonumber
\end{eqnarray}
where $\xi_z$ is the redshift evolution factor \citep{Murase:2016gly}, $f_{\rm bol}$ is the bolometric correction factor, and $f_{\gamma/p}=L_\gamma/L_p$. This estimate is consistent with the numerical result for $1-10$ GeV. The cutoff energy for gamma-rays due to the $\gamma\gamma$ pair production is $\sim2$ GeV for LERGs of $\dot{m}\sim\dot{m}_*$, so the gamma-ray spectrum has a peak around it.

The diffuse neutrino intensity has a peak at 10 PeV and is consistent with the extrapolation of the detected intensity \citep{ice15a,Aartsen:2015rwa}. It rapidly decreases above the energy due to the pion cooling suppression. The peak intensity of the neutrinos per flavor can be estimated to be $E_\nu^2\Phi_\nu\approx(1/3)f_{p\gamma}E_\gamma^2\Phi_\gamma\sim2\times10^{-9}\rm~GeV~s^{-1}~cm^{-2}~sr^{-1}$, where $f_{p\gamma}=t_{p\gamma}^{-1}/t_{\rm loss}^{-1}\sim0.1$ is the pion production efficiency at the peak energy. 
The predicted intensity is below the current upper limit provided by the extremely high-energy (EHE) analysis \citep{Aartsen:2018vtx}. The planned experiment, IceCube-Gen2 \citep{Aartsen:2019Gen2}, will be able to detect the predicted neutrinos, which provides a good test to our model.

We should note the uncertainty of the luminosity function of LERGs and the value of $\epsilon_j$. With the luminosity function given in Table 4 of \citet{2016MNRAS.460....2P}, the diffuse intensities are a factor of 3 lower. In this case, the contribution to the GeV gamma-ray background is less than 10 \% and detection of the neutrino background is challenging even with IceCube-Gen2. This uncertainty is shown by the shaded regions in Figure \ref{fig:diff}. On the other hand, the background intensities can be higher than those for our reference model for a lower value of $\epsilon_j$. The thin-lines in Figure \ref{fig:diff} represent the background intensities for $\epsilon_j=0.6$. MADs can provide $\sim50$ \% of the GeV gamma-ray background without significantly overshooting the current upper limit of the neutrino intensity. Future neutrino observations and accurate measurements of the radio-luminosity function will be able to clarify or constrain the parameter space of our MAD model.

Observationally, the gamma-ray luminosity in the LAT band correlates with the 5-GHz luminosity \citep{2011ApJ...733...66I,2014ApJ...780..161D,2019ApJ...879...68S}. Our model can reproduce the observed correlation at the lower luminosity range (bottom panel of Figure \ref{fig:diff}), taking into account the dispersion of $L_j$-$L_{\rm 1.4GHz}$ relation of 0.7 dex \citep{2010ApJ...720.1066C}. Here, we use $L_{\rm 5GHz}\approx(5/1.4)^{0.2}L_{\rm 1.4GHz}$ as in the previous works \citep{WRB01a,2011ApJ...733...66I}. For radio galaxies of $L_{\rm 5GHz}\gtrsim10^{40}\rm~erg~s^{-1}$, gamma-rays from MADs are attenuated by the $\gamma\gamma$ pair production (see Section \ref{sec:discussion} for its effects on maximum luminosity). Hence, our model predicts a sub-population that shows a fainter $L_\gamma$ than the relation, but detecting them by LAT is challenging. For $L_{\rm 5GHz}\gtrsim10^{43}\rm~erg~s^{-1}$, the assumption of the collisionless accretion flow breaks down, and jets should be responsible for the gamma-ray emission.

\section{Particle Injection at Magnetosphere }\label{sec:load}

The polar region above the central BH lacks mass supply because the centrifugal barrier and possible globally ordered magnetic fields prevent thermal particles from entering there. The density in the region continues to decrease, leading to formation of a magnetosphere \citep{2004ApJ...611..977M,2018ApJ...868..146N}. A vacuum gap may open in the magnetosphere when the number density is below the Goldreich-Julian density \citep{1969ApJ...157..869G}, 
\begin{equation}
n_{\rm GJ}=\frac{\Omega_F B_h}{2\pi ec}\approx\frac{B_h}{8\pi e R_G}\simeq5.6\times10^{-4}B_{h,3}M_9^{-1}\rm~cm^{-3},\label{eq:ngj}
\end{equation}
where $B_h\approx\Phi_{\rm MAD}/(2\pi{R}_G^2)\simeq1.1\times10^3\dot{m}_{-4}^{1/2}M_9^{-1/2}$ G is the magnetic field strength at the horizon, $\Phi_{\rm MAD}\sim50\sqrt{\dot{M}cR_G^2}$ is the magnetic flux for MADs \citep{TNM11a}, and $\Omega_F\approx c/(4R_G)$ is the angular velocity of the field lines with the efficient energy extraction \citep{BZ77a}.
The gap accelerates charged particles, which highly influences the structures of bulk flow at the launching point of the jets and high-energy emission in the magnetosphere \citep[see][]{2018Galax...6..116R}.

First, we examine whether the gap can be screened by the gamma-rays emitted from the MAD. The gamma-rays create electron-positron pairs in the magnetosphere through interaction with lower energy photons.
The gap could open at $\sim2R_G$ \citep{2016ApJ...818...50H,2017PhRvD..96l3006L,2020ApJ...895..121C,2020ApJ...902...80K}, and we hereafter consider the magnetosphere, or the funnel, of size $2R_G$. This is smaller than the emission region of the high-energy photons, and thus, we assume the isotropic photon distribution. 
Most of the observed radio galaxies have soft photon spectra in the GeV-TeV gamma-ray band. Then, the pair injection rate to the funnel is roughly estimated to be
\begin{eqnarray}
&\dot{N}_{\pm,\rm fun}&\approx \sigma_{\gamma\gamma}n_{\rm GeV}n_{\rm keV}c \mathcal{V}_{\rm fun}\sim \frac{\sigma_{\gamma\gamma}L_{\rm GeV}L_{\rm keV}}{16\pi^2R^4cE_{\rm GeV}E_{\rm keV}}\mathcal{V}_{\rm fun}\nonumber\\
&\simeq&2.5\times10^{38}M_9^{-1}\mathcal{R}_1^{-4}L_{\rm GeV,41}L_{\rm keV,40} \rm~s^{-1},\label{eq:Ndotpair}
\end{eqnarray}
where $n_{\rm GeV}=L_{\rm GeV}/(4\pi R^2cE_{\rm GeV})$ and $n_{\rm keV}$ are the photon number densities in the GeV and keV band, respectively, $L_{\rm GeV}$ and $L_{\rm keV}$ are the photon luminosities in the bands, $\mathcal{V}_{\rm fun}=4\pi(2R_G)^3/3$ is the volume of the magnetosphere. We use $\sigma_{\gamma\gamma}\approx0.2\sigma_T$ for the estimate, where $\sigma_T$ is the Thomson cross-section.
In the steady state, the pair production rate should be balanced by the advective escape rate, which leads to \citep[cf.][]{2011ApJ...730..123L}
\begin{eqnarray}
&n_\pm&\sim\frac{\dot{N}_{\pm,\rm fun}}{\mathcal{S}_{\rm fun}V_{\rm esc}}\label{eq:npair}\\
& \simeq&0.02M_9^{-3}\mathcal{R}_1^{-4}L_{\rm GeV,41}L_{\rm keV,40}\rm~cm^{-3},\nonumber
\end{eqnarray}
where $\mathcal{S}_{\rm fun}\approx4\pi(2R_G)^2$ is the surface area of the funnel and $V_{\rm esc}\approx c/3$ is the mean escape velocity of the pairs from the region. This value is two orders of magnitude higher than $n_{\rm GJ}$ in Equation (\ref{eq:ngj}), and thus, the vacuum gap is likely closed when MADs are the emission region of the observed GeV-TeV gamma-rays.

We also numerically estimate the pair density in the magnetosphere for M87 and NGC 315. The pair density is represented by
\begin{equation}
 n_\pm\approx2R_G\int{dE}_\gamma n_{E_\gamma}\int d\mathcal{E}_\gamma n_{\mathcal{E}_\gamma}\mathcal{K}(x),
\end{equation}
where  $n_{E_\gamma}=L_{E_\gamma}/(4\pi R^2cE_\gamma)$ is the differential number density of the photons, $L_{E_\gamma}$ is the differential photon luminosity, and $\mathcal{K}(x)=0.652\sigma_T\log(x)(x^2-1)x^{-3}H(x-1)$ and $x=E_\gamma\mathcal{E}_\gamma/(m_e^2c^4)$ \citep{cb90}.
We find that the pair amounts are sufficient for screening the gap in M87 and NGC315 (see Table \ref{tab:param}).

However, the amount of the pairs are insufficient to account for the observed features of the jet. Radio observations of M87 demand $U_\pm/U_B\gtrsim10^{-6}$, which leads to  
\begin{equation}
n_{\pm,\rm radio}\approx \frac{U_\pm}{\gamma_mm_ec^2}\gtrsim5B_2^2\gamma_{m,2}^{-1}\rm~cm^{-3},
\end{equation}
where $\gamma_m$ is the minimum Lorentz factor of the radio emitting pairs \citep{2015ApJ...803...30K}.  This value is around $10^5$ times higher than $n_{\rm GJ}$ for M87 given in Table \ref{tab:param}, which is a few orders of magnitude higher than our results. As another observed jet feature, the jet Lorentz factor at the parsec scale is estimated to be $\Gamma_j\sim10$ for typical AGNs including M87 \citep{2019ApJ...887..147P}. The required mass injection rate at the funnel region is then given by 
\begin{eqnarray}
&\dot{M}_{\rm fun}&\approx\frac{\epsilon_j\dot{M}}{(1+\sigma)\Gamma_j}\\
&\simeq&7\times10^{20}M_9\dot{m}_{-4}\epsilon_{j,0}\Gamma_{j,1}^{-1}\left(\frac{1+\sigma}{2}\right)^{-1}\rm~g~s^{-1}, \nonumber
\end{eqnarray}
where $\sigma$ is the magnetization parameter.
This is 9 orders of magnitude higher than that given by Equation (\ref{eq:Ndotpair}). Another mass injection mechanism is necessary to reproduce the observed jet features.

The pairs injected to the magnetosphere lose their energies by synchrotron radiation, and for typical parameters of $B_h\sim10^3$ G and $M\sim10^9~\msun$, all the pairs become $\gamma_\pm\sim1$ before their advective escape, where $\gamma_\pm$ is the Lorentz factor of the pairs. Typical energies of synchrotron emission is in X-rays to MeV gamma-rays, and these photons have too low energies to further produce the pairs by themselves. They may increase the pair production rate by acting as target photons for GeV gamma-rays. Here, we have ignored this effect to make a conservative estimate, but this effect will not change our conclusion.


Next, we examine mass and charge loading by neutrons. Neutrons are not deflected by the magnetic fields, so they can freely enter the magnetosphere. A neutron decays to a proton, an electron, and a neutrino in $\sim10^{3}(E_n/1~\rm GeV)$ sec, where $E_n$ is the neutron energy, and thus, they may inject baryons and charges into the funnel region. The distance between the neutron emission region and  the magnetosphere is approximated to be the size of the plasma, $R\sim 1.5\times10^{15}M_9\mathcal{R}_1$ cm. The mean free length of the neutron is written as $\sim3\times10^{13}(E_n/1~\rm GeV)$ cm. Hence, high energy neutrons of $E_n \sim30$ GeV provide a dominant contribution to the mass loading to the funnel. Then, the main channel of neutron production is the $pp$ inelastic collisions (see Figure \ref{fig:times}). The neutron loading rate is roughly estimated to be 
\begin{eqnarray}
& &\dot{N}_n\approx f_{pp} f_{\rm bol}f_{\rm geo}\frac{L_p}{E_{p,\rm cut}}\\
&\simeq& 1\times10^{33}f_{\rm bol,2}f_{pp,-3}f_{\rm geo,-2}.L_{p,42.5}E_{p,\rm cut,6.5}^{-1}\rm~s^{-1},\nonumber
\end{eqnarray}
where $f_{\rm bol}\approx(2-s_{\rm inj})(30\rm~GeV/E_{p,\rm cut})^{1-s_{\rm inj}}$ is the bolometric correction, $f_{pp}=t_{pp}^{-1}/t_{\rm loss}^{-1}$ is the $pp$ collision efficiency, and $f_{\rm geo}$ is the geometrical factor. The values of $f_{pp}$ and $f_{\rm bol}$ are based on a similar parameter set for M87 and NGC 315 (see Figure \ref{fig:times} for $f_{pp}$), and the value of $f_{\rm geo}$ is seen in Figure 2 in \citet{tt12}. The estimated neutron loading rate is several orders of magnitude lower than the pair injection rate. Therefore, neutrons do not contribute to either mass or charge loading in our model. We need other baryon loading mechanisms to make baryonic jets, or jets should mainly consist of pairs.

\section{Discussion}\label{sec:discussion}

Our parameter choice of high $\epsilon_{\rm NT}$ and low $s_{\rm inj}$ is based on particle-in-cell (PIC) simulations and theoretical considerations.   
Recent general relativistic MHD (GRMHD) simulations revealed that the accreting plasmas in the MAD regime have current sheets with $\sigma\gtrsim1$ \citep{BOP18a,2020ApJ...900..100R}. These simulations also indicate that guide fields are weak, compared to the reconnecting field, in a considerable fraction of the reconnection layers. In the relativistic reconnections with the magnetization parameter $\sigma\gtrsim1$, most of the dissipation energy is spent to produce non-thermal particles according to PIC simulations \citep{2014ApJ...783L..21S,2016ApJ...818L...9G,2018MNRAS.473.4840W,2018MNRAS.481.5687P}. These simulations support a high value of $\epsilon_{\rm NT}$ in the MADs, although further studies that connect the microscopic PIC simulations to macroscopic GRMHD simulations are necessary to obtain a more concrete conclusion.

The stochastic particle acceleration mechanism \citep{bld06,sp08} may also produce high-energy protons with a hard spectrum in our model, but an extremely optimistic parameter set is demanded, such as the diffusion coefficient with a Bohm-like energy dependence and the turbulence strength of $\eta<10$.  Although the non-thermal protons work as a coolant in MADs, they do not change the dynamical structure of the accretion flow as long as $\epsilon_{\rm NT}\lesssim0.5$ \citep{ktt14}. 

Recently, \citet{Murase:2019vdl} and \citet{2019PhRvD.100h3014K,Kimura:2020thg} discussed the hadronic gamma-ray and neutrino emissions from coronae in Seyfert galaxies and hot accretion flows in low-luminosity AGNs, respectively. These works focused on the SANE scenario where the magnetic field strength in the bulk of the flow is much weaker, compared to our MAD model, because the main targets of these studies are radio-quiet AGNs. In such a situation, particle acceleration is inefficient, and proton synchrotron emission is not effective. The different efficiency of non-thermal particle production may cause the different detection rate of gamma-rays from radio galaxies and radio-quiet AGNs. These two scenarios use different parameter sets assuming different magnetic field configuration, but parameters in Seyfert coronae and MADs are somewhat similar. If we consider only stochastic acceleration, coexistence of the two scenarios is unlikely, because it demands different power-law index of the turbulent power spectrum. On the other hand, the two scenarios may coexist if we consider efficient reconnection acceleration in MADs and assume that it is very inefficient in coronae in Seyfert galaxies and SANEs in low-luminosity AGNs.

  \begin{figure*}[t]
   \begin{center}
    \includegraphics[width=\linewidth]{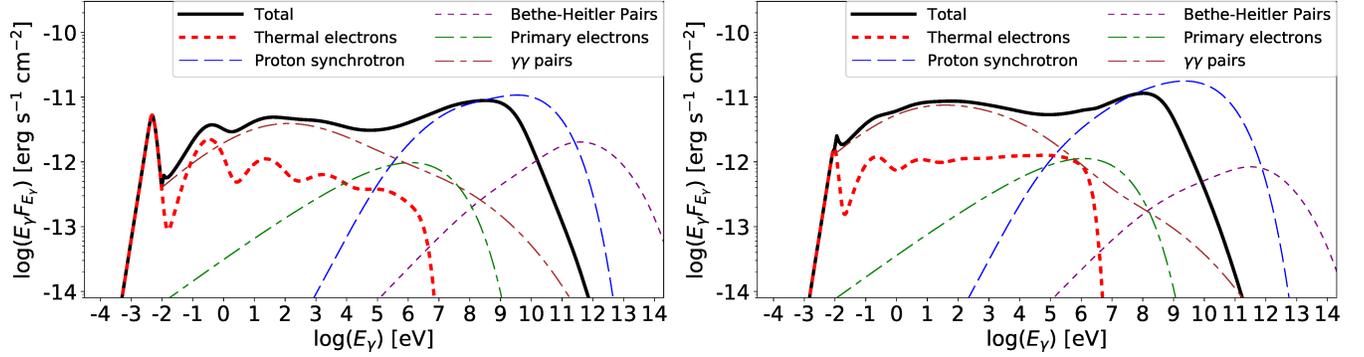}
    \caption{Same with Figure \ref{fig:spe}, but for the MADs with $\dot{m}=0.004$ (left) and $\dot{m}=0.02$ (right). }
    \label{fig:typical}
   \end{center}
  \end{figure*}

Our MAD model is applicable only to the hot accretion flows, which restrict the gamma-ray luminosity to be much lower than the Eddington luminosity. The critical accretion rate above which the hot accretion flows no longer exist is given by $\dot{m}_{\rm crit}\approx3\alpha^2\sim0.27$ for our reference parameter set \citep{1997ApJ...477..585M,xy12}. The proton luminosity for the MAD with $\dot{m}=\dot{m}_{\rm crit}$ is $L_{p,\rm crit}=\epsilon_{\rm NT}\epsilon_{\rm dis}\dot{m}L_{\rm Edd}\simeq2\times10^{44}M_9\alpha_{-0.5}^2\rm~erg~s^{-1}$. The gamma-ray luminosity is further lowered by the two-photon pair production for a high value of $\dot{m}$ due to copious X-ray photons emitted by the secondary electron-positron pairs by two-photon pair production. For a demonstration purpose, we calculate the broadband photon spectra for the cases with $\dot{m}=0.004$ and $\dot m=0.02$, i.e., $\dot m=\dot m_*$ and $\dot m=5\dot m_*$, and $d_L=100$ Mpc. The resulting spectra are shown in Figure \ref{fig:typical}. We can see that the gamma-ray spectra are strongly suppressed above $\sim1$ GeV (0.1 GeV) for $\dot m=0.004$ (0.02) due to $\gamma\gamma$ pair production. Because of the suppression, detection by {\it Fermi} is challenging for $\dot m\gtrsim0.01$. Then, the maximum gamma-ray luminosity by our MAD model in the {\it Fermi} band can be estimated to be $\sim6\times10^{43}M_9\rm~erg~s^{-1}$. Some radio galaxies have higher gamma-ray luminosities, and the gamma-ray production should be attributed to the jets. For high $\dot m$  cases, a relatively flat spectrum below 0.1 GeV will be detected by future MeV satellites, which will also provide additional tests of our MAD model. Also, for $\dot m\gtrsim0.01$, the photons produced by secondary pairs can work as target photons for hadronic interactions, though the $p\gamma$ cooling timescale is still lower than the synchrotron cooling timescale for $\dot m=0.02$. For a higher $\dot m$ case, non-linear electromagnetic cascades should be solved to calculate the complete spectra from MAD, which remains as a future work.

Neutrino luminosity from MADs are also limited by the energy budget, $L_\nu< L_{p,\rm crit}/2$. In MADs, neutrino luminosity is much lower than the proton luminosity because of the low pion production efficiency at low energies ($E_\nu<1$ PeV for $\dot{m}\sim\dot{m}_*$) and efficient meson coolings at high energies ($E_\nu>10$ PeV for $\dot{m}\sim\dot{m}_*$). Hence, MADs cannot be detected as a point neutrino source with the current neutrino facilities. Future experiments with a larger effective area may detect neutrinos from radio galaxies by stacking nearby luminous sources, which remains as a future work.  Note that the observed neutrino luminosity of TXS 0506+056, the blazar associated with IceCube neutrino events, is $L_\nu\sim10^{47}\rm~erg~s^{-1}$  \citep{Aartsen2018blazar1,Aartsen2018blazar2}, which is much higher than our theoretical upper limit of neutrino luminosity. The neutrinos from this source should be attributed to the jets directed to us \citep{2018ApJ...864...84K,2019NatAs...3...88G}.

Although our model is also applicable to BL Lac objects, a sub-class of blazars that have hot accretion flows around the BH, the high-energy emission from the jets completely dominates over that from the MADs due to the relativistic beaming effect. The intrinsic jet power launched from a MAD is comparable to the accretion power, $L_j\sim\epsilon_j\dot{M}c^2$ with $\epsilon_j\sim1$. For the steady state jets with $\theta_j>\Gamma_j^{-1}$, where $\theta_j$ is the jet opening angle and $\Gamma_j$ is the jet Lorentz factor, the observed radiation luminosity can be written as $L_{\rm obs}\approx(4/\theta_j^2) L_{\rm rad}$, where $L_{\rm rad}\approx \epsilon_{j,\rm NT}L_j$ is the jet radiation luminosity in the black-hole rest frame and $\epsilon_{j,\rm NT}$ is the conversion factor from the jet kinetic energy to the radiation \citep{1997ApJ...484..108S}.
Then, the ratio of the observed radiation luminosity from the jet to that from the MAD is estimated to be $\sim4\epsilon_{j,\rm NT}\epsilon_j\theta_j^{-2}/(\epsilon_{\rm NT}\epsilon_{\rm dis})\simeq100\theta_{j,-0.7}^{-2}(\epsilon_{j,\rm NT}\epsilon_j)/(\epsilon_{\rm NT}\epsilon_{\rm dis})$. Hence, radiation from the jets completely dominates over that from the MADs.  
On the other hand, such a luminosity enhancement is not expected for off-axis jets, and then, the emission from the MADs can be comparable to that from the jets as long as $\epsilon_{j,\rm NT}\epsilon_j\sim\epsilon_{\rm NT}\epsilon_{\rm dis}$.

Radio galaxies are also discussed as the origin of the ultrahigh-energy cosmic rays \citep[UHECRs; e.g.][]{Tak90a,MDT12a,RFG18a,2018JCAP...02..036E}. Although MADs can accelerate the protons up to $\sim1-10$ EeV, they mainly lose their energies by hadronic processes. Also, the accelerated particles in MADs are likely to be the same composition as the accreting plasma of the solar abundance, and accelerated ultrahigh-energy cosmic-ray nuclei are easily destroyed by the photo-disintegration process, leading to a proton-dominated composition. This is inconsistent with the Auger data that support a heavier composition \citep{AugerPAO14a}. Thus, the majority of the observed UHECRs should be accelerated in another cite \citep[e.g.,][]{Cap15a,KMZ18a}. For $\dot{m}\lesssim10^{-4}$, non-negligible fraction of protons can escape from the system, and may be observed on Earth. \citet{TMD16a} estimated the upper limit of the UHECR luminosity to be $\sim10^{42}\rm~erg~s^{-1}$ for the object of $d_L\sim40$ Mpc using the isotropic arrival direction of UHECRs. The UHECR luminosity of M87 in our model is lower, but the distance is closer.   It remains as a future work to investigate the possibility whether MADs in nearby radio galaxies contribute to the observed CRs for 1-10 EeV.  

\section{Summary}\label{sec:summary}

We propose hadronic interactions in MADs as a novel mechanism of the GeV-TeV gamma-ray production in radio galaxies. Non-thermal protons accelerated in MADs produce GeV gamma-rays through synchrotron radiations, and the secondary pairs produced by the Bethe-Heitler process emit TeV gamma-rays. These emissions can reproduce the gamma-ray data of M87 and NGC 315. 
Primary electrons co-accelerated with the protons emit MeV gamma-rays. M87 is detectable by the proposed MeV satellites, which will be useful to test the MAD model. Our MAD model produces relatively flat spectra for a broad range of photon energies, in contrast to the two-hump spectra often seen in the leptonic jet models. Hard X-ray and MeV gamma-ray observations are important to distinguish between these models.

The MADs can significantly contribute to the extragalactic gamma-ray background for $E_\gamma\sim1-30$ GeV up to 20 \% for our fiducial model. They also produce the extragalactic neutrino background of $E_\nu\sim10$ PeV, which is consistent with the current upper limit by IceCube. These neutrinos are detectable by IceCube-Gen2 and useful to test the particle acceleration process in MADs. Our model can reproduce the observed relation of radio and GeV gamma-ray luminosity for a low gamma-ray luminosity range. For luminous objects, GeV gamma-rays cannot escape from the MADs because of two-photon pair production, and thus, gamma-rays from luminous radio galaxies should be attributed to emission from their jets. Our model also predicts a population that is bright in radio but dim in GeV gamma-rays. These objects emit strong MeV gamma-rays, and future MeV satellites will be able to detect such a population.

We also evaluate the mass and charge loading to the black hole magnetosphere, or the funnel, by non-thermal processes in MADs. The gamma-rays from the MADs create significant amount of electron-positron pairs in the magnetosphere through two-photon pair production, which can screen the vacuum gap, prohibiting the particle acceleration there. Neutrons created in MADs are not injected to the funnel efficiently, due to the hard proton spectrum and low $pp$ collision efficiency. The amount of the pairs injected by two-photon pair production is not sufficient to explain the observed jet features. Further studies are necessary to understand the mass-loading mechanism to the jet-launching region.


\acknowledgements
We thank Shota Kisaka and Kohta Murase for fruitful discussion and Daniel Mazin for comments in the workshop "Active Galactic Nucleus Jets in the Event Horizon Telescope Era" at Tohoku University on Jan 20-22, 2020. We also thank the anonymous reviewer for comments that improve our manuscript.  This work is partly supported by JSPS Research Fellowship No. 19J00198 (S.S.K.) and KAKENHI No. 18H01245 (K.T.).

\bibliographystyle{aasjournal}
\bibliography{ssk}

\listofchanges

\end{document}